\documentclass{aastex61}

\usepackage{hyperref}
\newcommand{\kms}{$\rm ~km\,s^{-1}$}
\newcommand{\jyb}{$\rm Jy\,beam^{-1}$}
\newcommand{\hh}{$\rm H_2$}
\newcommand{\soa}{SO $6_5$--$5 _4$}
\newcommand{\sob}{SO $9_8$--$8 _7$}

\shorttitle{Molecular bullets in a high-mass star-forming region}
\shortauthors{Cheng et al.}

\begin{document}

\title{Multi-line observations of molecular bullets in a high-mass protostar}

\correspondingauthor{Keping Qiu}
\email{kpqiu@nju.edu.cn}

\author{Yu Cheng}
\affil{School of Astronomy and Space Science, Nanjing University, 163 Xianlin Avenue, Nanjing 210023, China}
\affil{Dept. of Astronomy, University of Virginia, Charlottesville, Virginia 22904, USA}
\affil{Dept. of Astronomy, University of Florida, Gainesville, Florida 32611, USA}

\author[0000-0002-5093-5088]{Keping Qiu}
\affil{School of Astronomy and Space Science, Nanjing University, 163 Xianlin Avenue, Nanjing 210023, China}
\affil{Key Laboratory of Modern Astronomy and Astrophysics (Nanjing University), Ministry of Education, Nanjing 210023, China}

\author{Qizhou Zhang}
\affiliation{Harvard-Smithsonian Center for Astrophysics, 60 Garden Street, Cambridge, MA 02138, U.S.A.}

\author{Friedrich Wyrowski}
\affiliation{Max-Planck-Institut f\"{u}r Radioastronomie, Auf dem H\"{u}gel 69, 53121 Bonn, Germany}

\author{Karl Menten}
\affiliation{Max-Planck-Institut f\"{u}r Radioastronomie, Auf dem H\"{u}gel 69, 53121 Bonn, Germany}

\author{Rolf G\"{u}sten}
\affiliation{Max-Planck-Institut f\"{u}r Radioastronomie, Auf dem H\"{u}gel 69, 53121 Bonn, Germany}

\begin{abstract}
We present Submillimeter Array (SMA) observations in the CO $J=3$--2, SiO $J=5$--4 and 8--7, and \sob\ lines, as well as Atacama Pathfinder EXperiment (APEX) observations in the CO $J=6$--5 line, of an extremely high-velocity and jet-like outflow in high-mass star-forming region HH 80--81. The outflow is known to contain two prominent molecular bullets, namely B1 and B2, discovered from our previous SMA CO $J=2$--1 observations. While B1 is detected in all the CO, SiO, and SO lines, B2 is only detected in CO lines. The CO 3--2/2--1 line ratio in B1 is clearly greater than that in B2. We perform a large velocity gradient analysis of the CO lines and derive a temperature of 70--210 K for B1 and 20--50 K for B2. Taking into account the differences in the velocity, distance from the central source, excitation conditions, and chemistry between the two bullets, we suggest that the bullets are better explained by direct ejections from the innermost vicinity of the central high-mass protostar, and that we are more likely observing the molecular component of a primary wind rather than entrained or swept-up material from the ambient cloud. These findings further support our previous suggestions that the molecular bullets indicate an episodic, disk-mediated accretion in the high-mass star formation process.

\end{abstract}

\keywords{ISM: individual objects (HH 80--81) --- ISM: jets and outflows --- stars: formation --- stars: massive}

\section{Introduction}\label{S:intro}
Jets and outflows are commonly observed in star-forming regions \citep{Lada85,Bachiller96,Richer00,Arce07,Frank14,Bally16}. Molecular outflows in low-mass protostars were first interpreted as ambient gas swept up by a wide-angle wind \citep{Shu91}, or by a collimated jet through turbulent layers \citep{Canto91,Stahler94} or bow shocks \citep{Raga93,Masson93}. These simplified models met with difficulties in fully characterizing the kinematics of low-mass outflows observed at high angular resolutions \citep{Arce02,Lee02}. Consequently, theoretical and numerical efforts have been made to develop more sophisticated models, e.g., the disk-wind and X-wind models \citep{Konigl00,Shu00}. In these models a magnetohydrodynamic (MHD) wind is centrifugally driven from a rotating disk around a protostar, and collimates into a jet. Such a wind, also referred as the primary wind, could sweep up ambient molecular gas into a large scale ($\sim0.1$--1~pc) outflow.  The observed range of collimation for molecular outflows could be understood as the consequence of the interaction between a laterally stratified X-wind and a toroid-like cloud core \citep{Li96,Shang06}, or as a manifestation of the radial structure of the disk magnetic field, which in turn determines the mass loading into a disk-wind \citep{Pudritz06,Pudritz07}. On the other hand, some MHD simulations of the cloud core collapse and protostar formation and evolution show that a low velocity ($\sim1$--10\kms), wide-angle outflow is directly ejected from the outer region of a circumstellar disk, whereas a high velocity ($\sim100$\kms), collimated jet is launched near the inner edge of the disk \citep[e.g.,][]{Banerjee06,Machida08}. This appears to be consistent with recent Atacama Large Millimeter/submillimeter Array (ALMA) observations which reveal a launching radius of sub-AU scales on the disk for a collimated molecular jet \citep{Lee17}, and larger launching radii up to a few 10~AU for wide-angle outflows \citep{Bjerkeli16,Tabone17}. Despite these advances, the detailed physical processes that drive jets and outflows (e.g., the magneto-centrifugal force or magnetic pressure from a magnetic tower) are still not well understood \citep[e.g.,][]{Hennebelle08}.   

Among the outflows associated with low-mass Class 0 protostars at the earliest evolutionary stages, a small group, characterized by a high degree of collimation and extremely high-velocity (EHV) emission in the CO lines, are of particular interests to our understanding of the outflow driving mechanism. These EHV components often have de-projected velocities of more than 100\kms~and carry momenta comparable with those of classical outflows. In particular, instead of being a continuous jet, the EHV component often appears as a collection of discrete clumps, which are often referred as molecular bullets \citep{Bachiller90b}. In the CO spectra these bullets appear as discrete secondary peaks distinguished from the smooth emission from the line wings. These features have been found in a handful of low-mass Class 0 sources, including L1448 \citep{Bachiller90b}, HH 7--11 \citep{Bachiller90a}, VLA 1623 \citep{Andre90}, IRAS 03282+3035 \citep{Bachiller91}, IRAS 20050+2720 \citep{Bachiller95}, Cep E \citep{Lefloch96}, HH 111 \citep{Cernicharo96}, HH 211 \citep{Gueth99}, and IRAS 04166+2706 \citep{Tafalla04}. The bullets tend to be regularly spaced along the axis of a molecular jet and often show symmetry on both sides of the driving source,  and coincide with shocked \hh~ knots and optical emissions in some sources \citep[e.g.,][]{Hatchell99}. 

Different from classical molecular outflows, these clumpy and EHV jets seem to represent a special outflow component with a unique origin \citep{Bachiller96}. Various models have been proposed to explain how the bullets are produced, however, none of them appears to be capable of explaining all the sources. \citet{Masson90} suggested that the EHV component in HH 7--11 could be the gaseous material swept up by an underlying jet interacting with the ambient cloud. However, such a strong interaction would create J-type shocks that may completely destroy molecular material \citep{Hollenbach77}. \citet{Hatchell99} argued that the bullets in HH 111 and Cep E cannot consist of a swept-up material given their high masses. Other models proposed to interpret molecular bullets include: 1) the ``bullet'' model, i.e., direct ejections from the close vicinity of a central protostar \citep[e.g.,][]{Bachiller90b} and 2) the internal working surface (IWS) model \citep[e.g.,][]{Raga92,Santiago09}. In the former scenario molecular bullets represent an impulse enhancement in mass ejection powered by the central star-disk system, and are probably the precursors of Herbig-Haro (HH) objects, as described by the interstellar bullet model \citep{Norman79}. In the latter model, supersonic velocity variations of an underlying jet induce IWSs of internal shocks, which appear to be bullet-like structures. Both models could explain the symmetric appearance of bullets since the tight coupling across the thickness of a protostellar disk leads to simultaneous fluctuations of mass ejections in the opposite directions.

Outflows in high-mass protostars, since being first discovered in 1970s \citep{Zuckerman76}, have attracted great interests and motivated both single-dish surveys and interferometer case studies. The outflow energetics appear to correlate with the source luminosity up to $10^6~L_{\odot}$ \citep{Shepherd96,Beuther02,Wu04,Zhang05,Maud15}. This correlation gives rise to an inspiring question whether outflows in the full mass spectrum share a common driving mechanism. Wide-angle outflows in high-mass star-forming regions can have a gas mass reaching $100~M_{\odot}$ \citep[e.g.,][]{Shepherd98,Qiu09b}, which is an order of magnitude larger than the mass of the central forming star. It seems implausible that such massive outflows arise from materials ejected from a circumstellar disk, and rather a good fraction of the outflow gas may come from the ambient cloud. Indeed, there is growing evidence from observations that a wind-ambient cloud interaction similar to what is proposed for low-mass outflows may occur in the high-mass regime \citep{Qiu09b,Li13,Liu18}. Theoretically, there have been several numerical studies finding that an MHD wind from the circumstellar disk could be the driving mechanism of massive outflows \citep[e.g.,][]{Matsushita18}. 

Furthermore, existing observations of EHV molecular jets and bullets are mostly limited to low-mass outflows. \citet{Qiu09a} reported the first detection of molecular bullets in high-mass star-forming region HH 80--81, which has a bolometric luminosity of 2$\times$10$^4$ $L_\odot$ at an adopted distance of 1.7 kpc \citep{Rodriguez80}. This region is best known for a spectacular (10.3 pc long) radio jet, which is launched from a high-mass star-forming core, namely MM1, and powered by an early B-type protostar \citep{Marti93,Marti95,Marti98,Qiu09a,Masque12,Masque15,Qiu19}. Another molecular cloud core, MM2, with a mass $>12~M_\odot$, is separated by about 7\arcsec~from MM1 and appears to be at an earlier evolutionary stage \citep{Qiu09a,Fernandez-Lopez11}. At least two molecular bullets are seen in an EHV jet-like outflow emanating from MM2 \citep{Qiu09a}. The bullets were identified in the CO $J=2$--1 velocity integrated maps as well as in the mass-velocity and position-velocity diagrams. These bullets exhibit similar appearance as their counterparts in low-mass outflows, and were suggested to originate from the close vicinity of the central high-mass protostar. While the findings are intriguing, other explanations could not be ruled out based on their observations. Here we present new observations of this region, including data in the CO $J=3$--2 and 6--5, SiO $J=5$--4 and 8--7, and \sob~ lines to further investigate the origin of the molecular bullets. We describe the observations and data reduction in Section \ref{S:obs}. Results and analyses are presented in Sections \ref{S:res} and \ref{S:lvg}, respectively, followed by discussions in Section \ref{S:dis}. A brief summary is presented in Section \ref{S:sum}.

\section{Observations}\label{S:obs}
The Submillimeter Array\footnote{The SMA is joint project between the Smithsonian Astrophysical Observatory and the Academia Sinica Institute of Astronomy and Astrophysics and is funded by the Smithsonian Institution and the Academia Sinica.} (SMA) observations were performed toward (R.A., Decl.)$_{\rm J2000}$=($18^{\rm h}19^{\rm m}12.\!^{\rm s}5$, $-20^{\circ}47'27''$) in two epochs, and covered the CO 3--2 line at 345.796 GHz, the SiO 5--4 line at 217.105 GHz, the SiO 8--7 line at 347.331 GHz, and the \sob~line at 346.528 GHz. The observations in the SiO 5--4 line were carried out on 2009 April 14, when the SMA was in its Compact configuration with 6 available antennas. The weather was excellent with the zenith atmospherical opacity at 225 GHz, $\tau_{\rm 225GHz}$, around 0.07 during the observations. The data were taken with the 230~GHz receivers and processed by the correlators covering 213.5--217.5 GHz in the lower sideband (LSB) and 225.5--229.5 GHz in the upper sideband (USB), with a uniform spectral resolution of 406.25 kHz ($\sim$0.56\kms). The observations in the CO 3--2, SiO 8--7, and \sob~lines were performed in the Compact configuration with 6 available antennas on 2010 April 9 under an excellent weather condition  with $\tau_{\rm 225GHz}\sim$0.07-0.08. The data were taken with the 345~GHz receivers and processed by the correlators covering 333.6--337.6 GHz in the LSB and 345.6--349.6 GHz in the USB, with a uniform spectral resolution of 812.5 kHz ($\sim$0.70\kms). For both observations, we observed 3C273 for bandpass calibrations, Titan for absolute flux calibrations, and two quasars, J1733-130 and J1911-201, for time dependent gain calibrations. The data were calibrated using the IDL MIR package\footnote{https://github.com/qi-molecules/sma-mir}, and the calibrated data were then exported to MIRIAD \citep{Sault95} for imaging. We make final data cubes of all the spectral lines with a velocity resolution of 1.2\kms, to match the velocity resolution of our previously observed CO~2--1 and \soa~lines \citep{Qiu09a}. The CO~3--2, SiO~8--7, and \sob~maps have synthesized beams of $3.\!''1\times1.\!''8$ with position angles (PAs) of $-8^{\circ}$ at the full width half maximum (FWHM), and root mean square (RMS) sensitivities of approximately 60~m\jyb\ per velocity channel. The SiO~5--4 map has a synthesized beam of $4.\!''1\times3.\!''3$ with a PA of $-43^{\circ}$ at FWHM, and an RMS sensitivity of approximately 20~m\jyb. We convolve the SiO~8--7 map to the angular resolution of the SiO~5--4 map, and also convolve the \sob~map to a beam of $3.\!''7\times3.\!''0$ with a PA of $-7^{\circ}$, which is the synthesized beam at FWHM of our previous \soa~map \citep{Qiu09a}. We retain the original angular resolution of the CO~3--2 data when investigating the bullet morphology, but convolve the map to the beam of our previous CO~2--1 data \citep{Qiu09a} for measuring the CO~3--2/2--1 line ratios and for subsequent analysis.

The Atacama Pathfinder EXperiment\footnote{This publication is based on data acquired with the Atacama Pathfinder Experiment (APEX). APEX is a collaboration between the Max-Planck-Institut fur Radioastronomie, the European Southern Observatory, and the Onsala Space Observatory.} (APEX) CO 6--5 observations were carried out with the Carbon Heterodyne Array of the MPIfR \citep[CHAMP$^+$][]{Kasemann06}. CHAMP$^+$ is a dual-color receiver array operating in the 450 and 350 $\mu$m atmospheric windows, allowing us to simultaneously observe the CO 6--5 and 7--6 lines. We only detected the CO 6--5 emission arising from the bullets. Therefore there is no further presentation or discussion on the CO 7--6 line in this work. The data presented here come from the same observing program as those presented in \citet{Qiu19}. Readers are referred to \citet{Qiu19} for a detailed description of the observations and relevant parameters.

\section{Results}\label{S:res}

\subsection{SMA observations}
\begin{figure*}
\epsscale{1.}\plotone{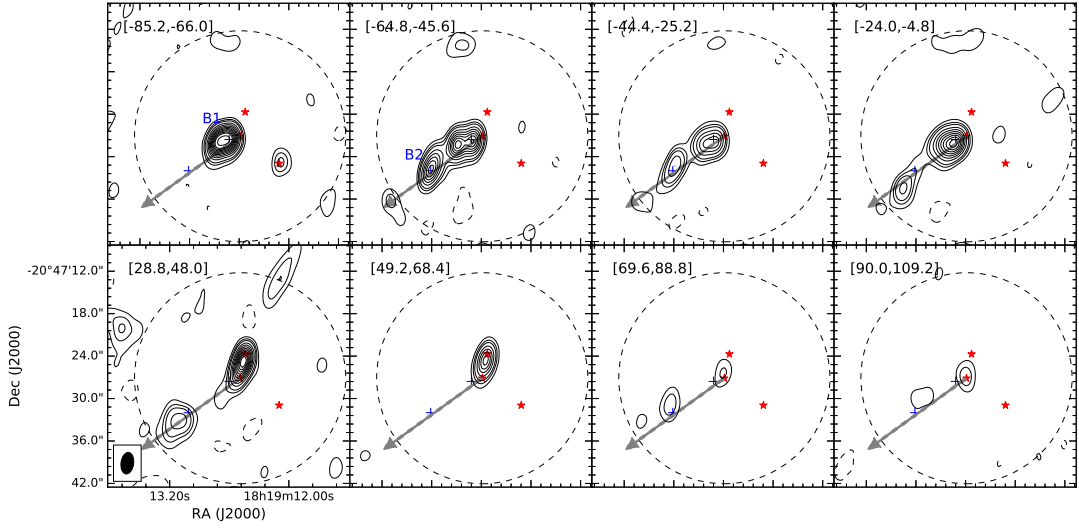}
\caption{Contour maps of the CO 3--2 emission integrated over every 20.4\kms, as indicated in the upper left of each panel, from $-$85.2 to $-$4.8\kms~and from 28.8 to 109.2\kms. The contour levels are ($-$1, 1, 2, 3, 5, 7, 9, 11, 13, 15, 17, 19, 21, 25, 30, 35, 40)$\times \sigma$ where $\sigma=$ 1.2, 1.8, 2.4, 3.0~\jyb\kms~for the upper panels from left to right and $\sigma=$ 3.0, 2.4, 1.8, 1.2~\jyb\kms~ for the lower panels from left to right, respectively. A grey arrow delineates the orientation of the SE outflow arising from MM2. The dashed circle marks the SMA primary beam. Three stars indicate dense cores MM1, MM2,  and MC \citep{Qiu09a}. Two blue pluses mark the peaks of the two molecular bullets B1 and B2. Hereafter, a filler or open ellipse located in the lower left corner of a panel shows the synthesized beam at FWHM.}
\label{fig:CO}
\end{figure*}

\begin{figure*}
\epsscale{1.}\plotone{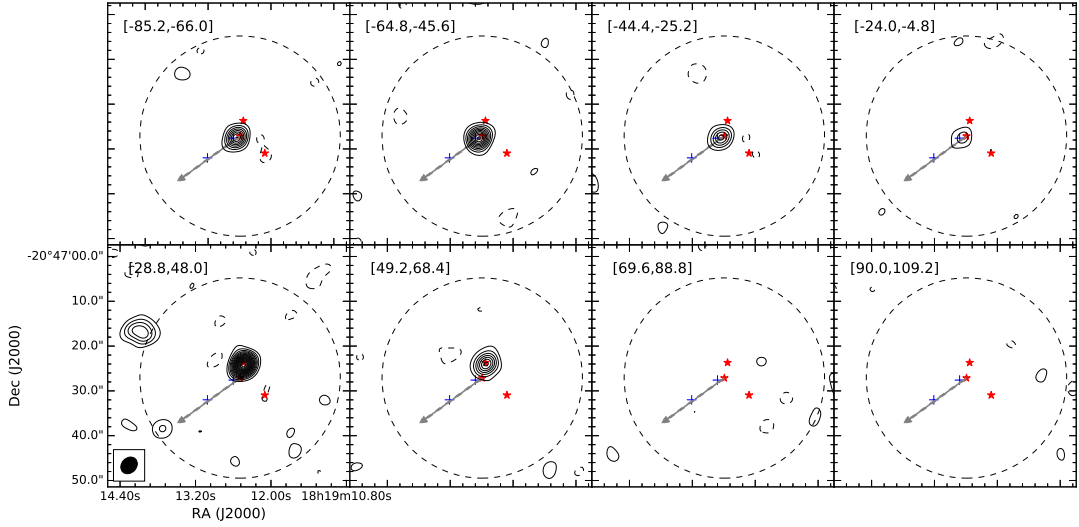}
\caption{Contour maps of the SiO 5--4 emission integrated over the same velocity ranges as those in Figure~\ref{fig:CO}; the first and spacing contour levels are 0.3~\jyb\kms. Other symbols are the same as those in Figure~\ref{fig:CO}.}
\label{fig:SiO}
\end{figure*}

Figure~\ref{fig:CO} shows the high-velocity CO 3--2 emission integrated over every 20.4\kms, from $-$85.2 to $-$4.8\kms~for the blueshifted lobe and from 28.8 to 109.2\kms~for the redshifted lobe, respectively. The CO 3--2 map can be directly compared with the CO 2--1 map in \citet{Qiu09a}. Similar to the CO 2--1 data, the CO 3--2 emission mainly traces a knotty and jet-like outflow originating from a dust peak, namely MM2 following  \citet{Qiu09a}, and pointing to the southeast (hereafter the SE outflow). The SE outflow is seen in both blueshifted and redshifted emissions, but appears much more remarkable in the blueshifted lobe (up to a velocity of $-$85.2\kms, corresponding to $|{\Delta}V|=97.2$\kms~with respective to the systemic velocity of 12\kms~for MM2). In particular, two EHV bullets, namely B1 and B2 following \citet{Qiu09a}, can be readily identified as compact knots in Figure~\ref{fig:CO}. The peak positions of each bullet derived from the CO 2--1 and 3--2 maps agree with each other within $0.\!''1$, indicating that the two lines are essentially tracing the same gas. Since the CO 2--1 and 3--2 data reveal essentially the same position, size, and velocity range of the bullets and the former have a higher signal-to-noise ratio (SNR), we list the parameters measured from the CO 2--1 data in Table \ref{tab:measured}. Figure~\ref{fig:SiO} shows the SiO 5--4 velocity channel maps. Different from the CO lines, the SiO 5--4 emission mostly traces structures close to MM2. The blueshifted emission arises from a clump with a broad velocity coverage of $-85.2$ to $-4.8$\kms, and shows no further elongation along the outflow axis. In the redshifted lobe, the SiO emission traces another compact clump associated with another dense core, namely MC following \citet{Qiu09a}, and that clump is also seen in the high-velocity emissions in the CO 3--2 and 2--1 line. 

\begin{figure*}
\epsscale{1.}\plotone{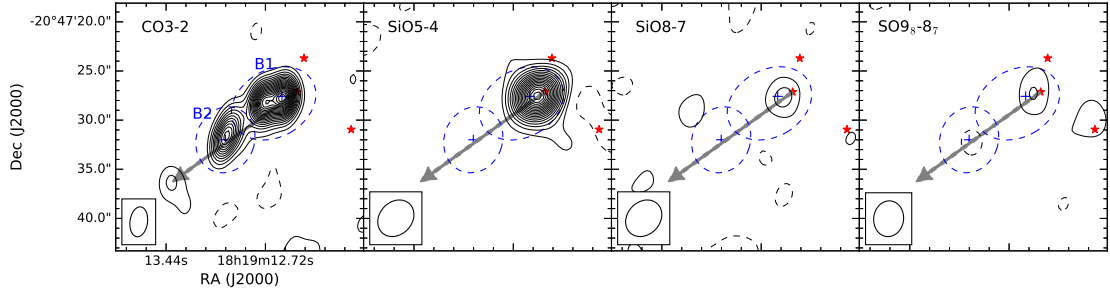}
\caption{From left to right, CO 3--2, SiO 5--4, SiO 8--7, and \sob~emissions integrated from $-$74.4\kms~to $-$34.8\kms. The starting and spacing contour levels are 3.0, 0.3, 1.3, 1.3~\jyb\kms~from left to right, respectively. Dashed ellipses delineate the positions and sizes of the two bullets. Other symbols are the same as those in Figure~\ref{fig:CO}.}
\label{fig:int}
\end{figure*}

\begin{deluxetable}{ccccc}
\tablecaption{Positions, sizes, and velocities of the bullets \label{tab:measured}} 
\tablehead{ 
\colhead{Source} &
\colhead{$\Delta \alpha$\tablenotemark{a}} &
\colhead{$\Delta \delta$\tablenotemark{a}} &
\colhead{Size\tablenotemark{b}} &
\colhead{Velocity\tablenotemark{c}} \\
\colhead{} &
\colhead{(arcsec)} &
\colhead{(arcsec)} &
\colhead{(pc)} &
\colhead{(\kms)}
}
\startdata B1 & 1.64 & $-0.45$ & 0.016 & [$-74.4$, $-57.6$]  \\
                B2 & 7.79 & $-4.87$ & 0.013 & [$-56.4$, $-34.8$]  \\
\enddata
\tablenotetext{a}{relative to the peak position of MM2: ($\rm 18^h19^m12\fs49$, $-20\arcdeg47\arcmin 27\farcs13$);}
\tablenotetext{b}{geometric mean of the de-convolved major and minor axes;}
\tablenotetext{c}{velocity ranges where the EHV bullet features dominate the line wing emission, see Section \ref{S:lvg}.}
\end{deluxetable}

In Figure~\ref{fig:int} we present the CO, SiO, and SO maps integrated from $-$74.4\kms~to $-$34.8\kms, the velocity range where the two bullets dominate the emissions. The dashed ellipses approximately delineate the locations and sizes of the CO bullets. Since the bullets appear distinct in both space and velocity, we can distinguish the emissions between B1 and B2 even though the two bullets are adjacent to each other in the CO 3--2 map. Bullet B1 is located about  2\arcsec~ to the southeast of  MM2, the driving source of the SE outflow and the bullets. Bullet B2 is located to the downstream of B1 and moves with a relatively lower velocity. There is a faint clump about 6\arcsec~ to the southeast of B2 along the jet path. This feature is also seen in the CO 2--1 map \citep{Qiu09a}, but was not identified as a bullet due to its weak emission and lack of a secondary peak in the CO spectra. Here we focus on bullets B1 and B2. B1 appears remarkably bright in the SiO 5--4 emission, and is also detected in the SiO 8--7 and SO $8_7$--$7_6$ lines. Note that the SO $6_5$--$5_4$ emission was also detected toward B1 as reported in \citet{Qiu09a}. However, there is no emission in any of the SiO or SO lines toward B2. The peak position of B1 in the SiO and SO maps is about 1\arcsec closer to MM2 as compared with that in the CO maps. Such a deviation might manifest differences in molecule formation mechanisms and/or excitation conditions. In a prototype of low-mass Class 0 outflows, HH 211, \citet{Lee10} found that for most of the spatially resolved knots, the CO peaks are slightly ahead of the SiO and SO peaks, possibly tracing different shock regions.

\subsection{APEX observations}
\begin{figure}
\epsscale{1.}\plotone{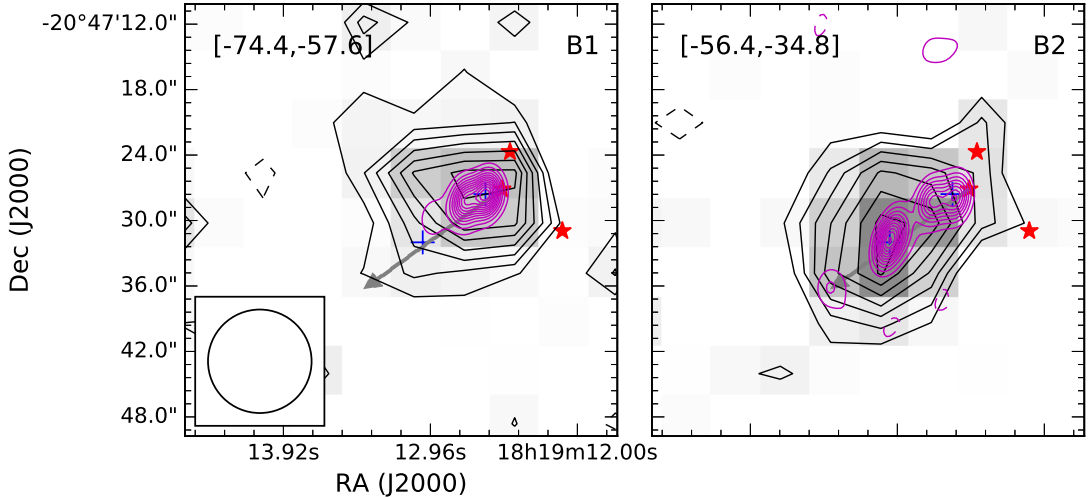}
\caption{
{\it Left}: CO 6--5 emission of bullet B1 integrated from $-74.4$ to $-57.6$\kms~shown in grey scale and contours, with contour levels starting at 2.0 K\kms~and continuing in steps of 1.0~K\kms~. Overlaid magenta contours indicate the SMA map of the CO 3--2 emission integrated over the same velocity range. The starting and spacing contour levels are 6.0~\jyb\kms. An open circle in the lower left shows the APEX beam at 691~GHz. {\it Right}: Same as the left panel, but for bullet B2, with the emission integrated from $-56.4$ to $-34.8$\kms. The CO 6--5 contour levels start at 3.2~K\kms~and continue in steps of 1.6~K\kms, and the CO 3--2 starting and spacing contour levels are 6.0~\jyb\kms. 
}\label{fig:CO65}
\end{figure}

\begin{figure}
\epsscale{1.}\plotone{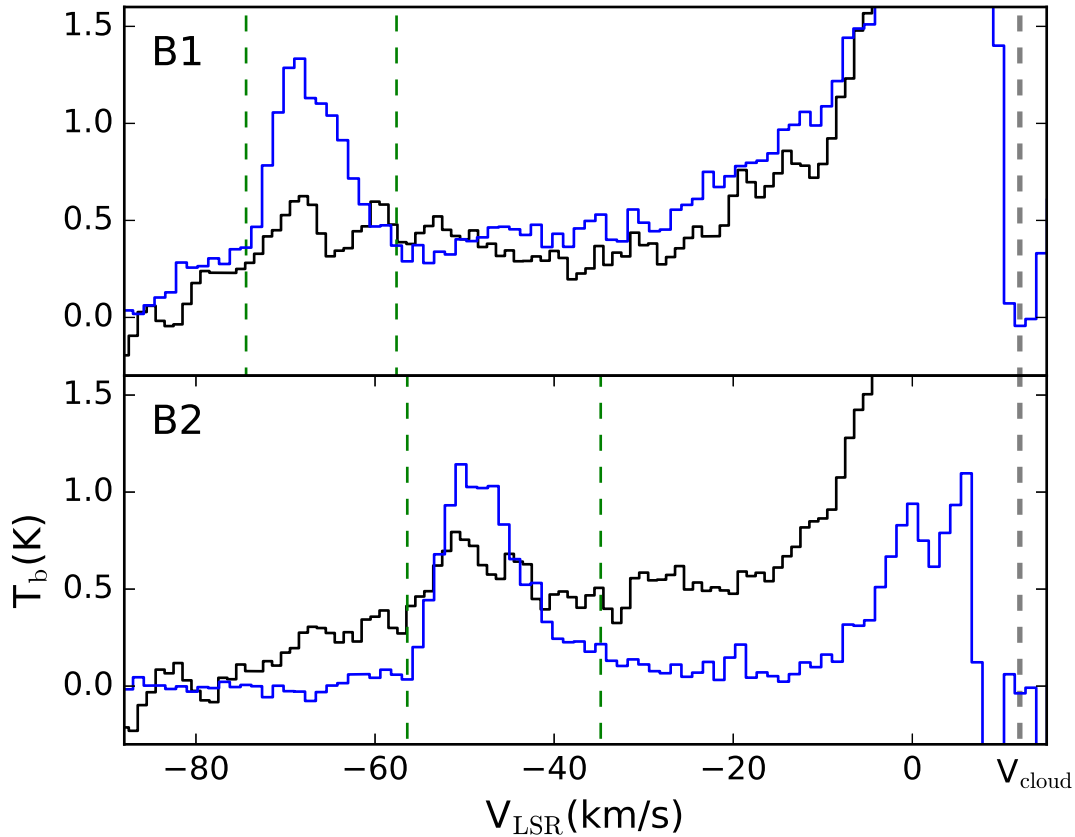}
\caption{CO 6--5 spectra of bullets B1 (upper panel) and B2 (lower panel), shown in black lines, taken from the positions indicated by the blue pluses in Figure~\ref{fig:CO}. The CO 3--2 spectra, taken from the same positions, are overlaid in blue lines. Note that for comparison the CO 3--2 amplitudes have been reduced by a factor 5. Vertical dashed green lines indicate the velocity ranges of the two bullets, and a grey line denotes the systemic velocity at about 12\kms.}
\label{fig:65spec}
\end{figure}

The CO 6--5 line has a critical density of order 10$^{\rm 5}$~cm$^{\rm -3}$ and $E_{\rm up}/k$ reaching 116~K, where $E_{\rm up}$ is the energy of the upper level above the ground and $k$ is the Boltzmann constant. This high critical density and high energy level transition probes the dense and warm gas. Figure~\ref{fig:CO65} shows the CO 6--5 emission integrated over two velocity ranges where the two bullets dominate. For comparison we also show the CO 3--2 emission integrated over the same velocity ranges. It is clear that the two bullets are both detected in the CO 6--5 emission, though the $9.\!''5$ beam of the APEX cannot fully resolve the bullets. Such high excitation CO lines are rarely seen in EHV outflows  \citep[see also][]{Leurini09}. Figure~\ref{fig:65spec} shows the APEX CO 6--5 spectra of the two bullets taken at the emission peaks, as well as the SMA CO 3--2 spectra taken at the same positions. The spectra are smoothed to a velocity resolution of 3.0\kms~to increase the SNR. The detection of the bullets in the CO 6--5 spectra is appreciable, but appears less prominent than that in the CO 3--2 spectra, probably due to a combination of the beam dilution effect and lower SNRs. There are hints in the CO 6--5 spectrum of B1 for multiple velocity components inside the bullet.  

\section{LVG ANALYSIS}\label{S:lvg}

\begin{figure}[t]
\epsscale{.8}\plotone{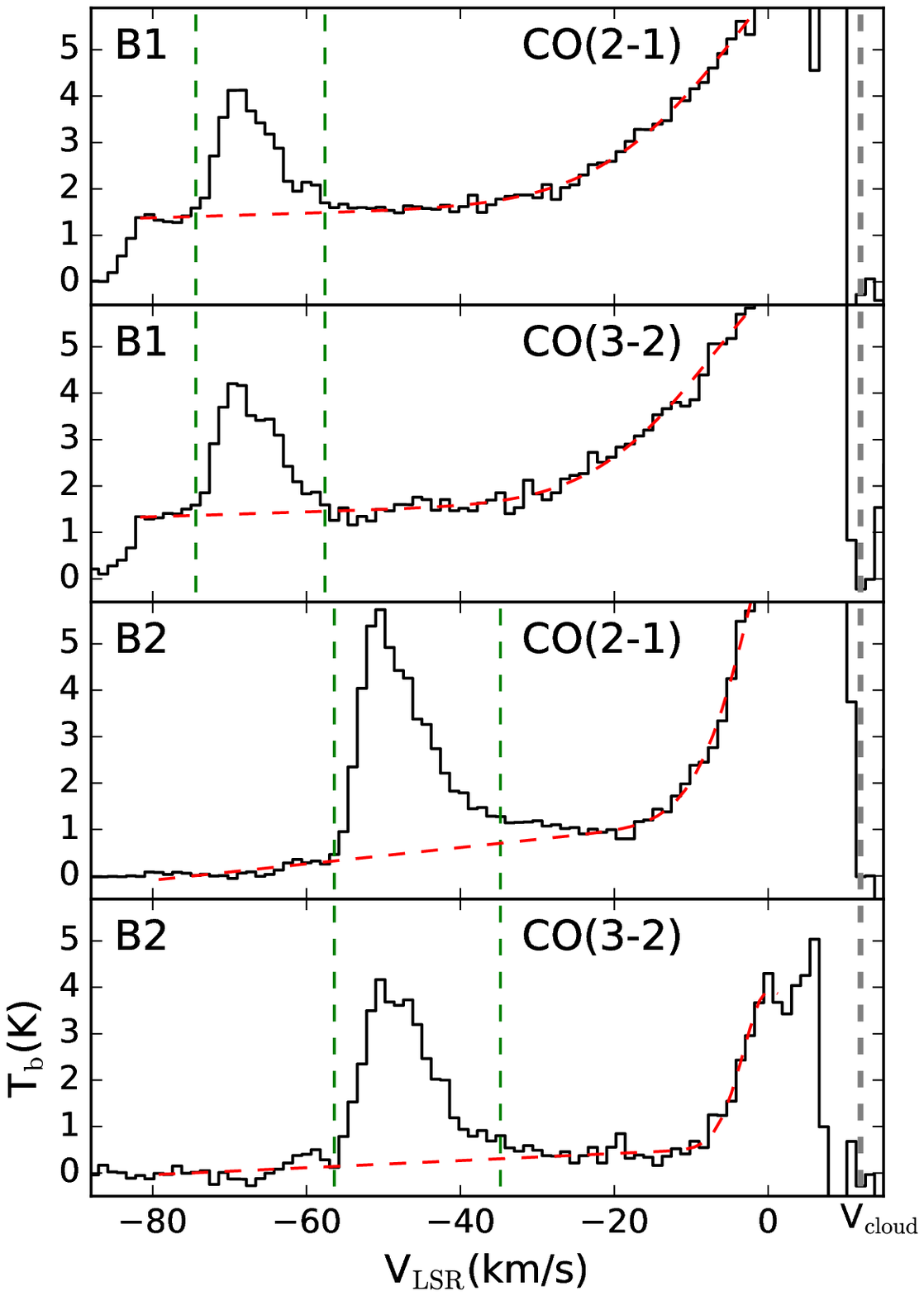}
\caption{SMA CO 2--1 and CO 3--2 spectra extracted from the peak positions of B1 and B2. In each panel, a dashed red line indicates the fitted broad line wing component (see the text in Section \ref{S:lvg} for more details). Other symbols are the same as those in Figure~\ref{fig:65spec}.}
\label{fig:cospec}
\end{figure}

To infer the excitation conditions of the two bullets, we performed a Large Velocity Gradient (LVG) analysis of the new CO 3--2 observations and the published CO 2--1 data \citep{Qiu09a}. The basic requisite for the LVG approximation to be valid is that, over a characteristic distance 
where physical conditions vary, the velocity shift arising from the gas kinematics or turbulence should be larger than the local thermal linewidth so that emitted photons corresponding to a certain spectral line can only be absorbed locally or escape to infinity. Such an approximation is naturally applicable to high velocity outflows, which have a broad velocity range owing to their macroscopic velocity structures \citep[see, e.g., detailed shock models in][]{Gusdorf08}. The escape probability of a spectral line photon in a given direction can be derived analytically as a function of optical depth for a given geometry, and thus LVG calculations of the observations allow to constrain physical conditions (density and temperature) and optical depth. The APEX CO 6--5 data are not included in the LVG nanlysis due to the lower spatial resolution and lower SNR. In addition, the coarse spatial resolution and an unknown pointing error of the CO 6--5 observations would induce contamination to the bullets flux by the material around the bullets, and thus significantly affect the reliability of the LVG calculations.

We convolve the CO 3--2 map to the resolution of  the CO 2--1 data. Figure~\ref{fig:cospec} shows the SMA CO 3--2 and 2--1 spectra of the two bullets. To measure the fluxes of the two bullets we inspect the spectra and channel maps, and identify the velocity ranges of the two bullets as $-$74.4\kms~to $-$57.6\kms~for B1 and from $-$56.4\kms~to $-$34.8\kms~for B2. In these two velocity ranges, the bullets clearly dominate the emission in both the spectra and velocity channel maps. In Figure~\ref{fig:cospec}, the EHV emission from each bullet in the CO 2--1 or 3--2 lines is remarkable, but appears as an abrupt peak superposed on a broad line wing component which starts at low velocities close to the cloud velocity and extends to the highest velocities. Therefore, it is not straightforward to accurately measure the CO emission fluxes of the bullets. In principle, one should subtract the wing component before measuring the fluxes of the bullets. To quantify the emission flux from the broad line wing within the bullet velocity ranges, we fit the component with various functions, and find that a combination of a Gaussian and a polynomial worked best, as shown in the red dashed lines in Figure~\ref{fig:cospec}. However, the fitting results largely depend on the selected velocity ranges, and it is difficult to assess the accuracy of the estimate of the broad line wing fluxes. The measured CO 2--1 and 3--2 fluxes, along with the CO 3--2/2--1 ratios, are given in Table \ref{tab:derived}. The flux uncertainties are mainly due to a flux calibration uncertainty of 10\%. In addition, the spatial filtering effect due to inadequate $(u,v)$ sampling in interferometer observations will introduce a flux loss from extended emissions, but such an effect should not be significant for compact and EHV emissions \citep[e.g.,][]{Qiu09b}, and could be further mitigated in the estimate of the line ratio. 

\begin{deluxetable*}{cccccc}
\tablecaption{Measured and derived properties of the bullets\label{tab:derived}} 
\tablehead{ 
\colhead{Source} &
\colhead{$\int{T_{\rm b}}$(CO~2--1)\,$dV$~\tablenotemark{a}} &
\colhead{$\int{T_{\rm b}}$(CO~3--2)\,$dV$~\tablenotemark{a}} &
\colhead{CO 3--2/2--1~\tablenotemark{a}} &
\colhead{Temperature~\tablenotemark{b}} &
\colhead{\hh~Density~\tablenotemark{b}} \\
\colhead{} &
\colhead{(K\kms)} &
\colhead{(K\kms)} &
\colhead{} &
\colhead{(K)} &
\colhead{($10^3$~cm$^{-3}$)} 
}
\startdata 
B1 & $47.4\pm4.8$ ($23.1\pm2.6$) & $48.1\pm4.8$ ($24.4\pm2.8$) & $1.02\pm0.14$ ($1.06\pm0.17$) & $123_{-54}^{+91}$ & $4.9_{-0.6}^{+0.7}$ \\
B2 & $61.8\pm6.2$ ($51.3\pm5.7$) & $45.1\pm4.5$ ($40.5\pm4.6$) & $0.73\pm0.10$ ($0.79\pm0.13$) & $32_{-12}^{+17}$ & $8.8_{-1.6}^{+3.2}$ \\
\enddata
\tablenotetext{a}{Measured toward the peak positions, and the values in parentheses are for the case with the wing component subtracted;}
\tablenotetext{b}{Calculated for the case without subtracting the line wing component.}
\end{deluxetable*}

We perform LVG calculations using the RADEX code \citep{vdTak07} to compute the brightness temperature of each CO line for given gas temperature ($T$), \hh~number density ($n$), and the CO column density-to-line width ratio ($N_{\rm CO}/{\Delta}V$). We adopt a typical CO-to-\hh~abundance ratio of $10^{-4}$, and assume that the dimension of the bullets along the line of sight is similar to that in the plane of sky. This way a CO column density can be determined from $n$ with a size scale measurable in the CO emission map (see Table~\ref{tab:measured}). We also obtain a line width of 8.0\kms~for B1 and 10.5\kms~for B2 (from Figure~\ref{fig:cospec}). We then generate a series of LVG models to fit the CO 2--1 and 3--2 fluxes, or equivalently the CO 2--1 flux and the CO 3--2/2--1 ratio, by varying $T$ and $n$  (the latter also determines $N_{\rm CO}/{\Delta}V$). To check for any bias that could be introduced in the fitting of the broad line wing, we perform LVG calculations using fluxes of the bullets measured both with and without subtracting the line wing component. It can be seen that whether subtracting the broad line wing component considerably affects the LVG model results for B1: we found $T\sim120$~K and $n\sim5\times10^3$~cm$^{-3}$ without line wing subtraction, and $T$ of 500~K and $n\sim10^3$~cm$^{-3}$ with the wing component subtraction. On the other hand, for B2, the gas temperature and density do not change much with and without line wing subtraction, with $T<50$~K and $n\lesssim9\times10^3$~cm$^{-3}$. This is not surprising since the integrated flux of the wing component constitutes more than 50\% of the total flux within the B1 velocity range, but is only 15\% of the total flux within the B2 velocity range (see Figure \ref{fig:cospec}). For the case that the line wing component is subtracted, B1 has a gas density of only $10^3$~cm$^{-3}$ and is almost an order of magnitude lower than that of B2. This does not seem to be plausible considering that the high excitation SO and SiO lines are detected in B1 but not in B2 (see Figure \ref{fig:int}). We will discuss the LVG model results in more details in Section \ref{subS:excitation}. Here in Table \ref{tab:derived} we only list the results for the case that the line wing component is not subtracted.

\begin{figure*}
\epsscale{1.}\plotone{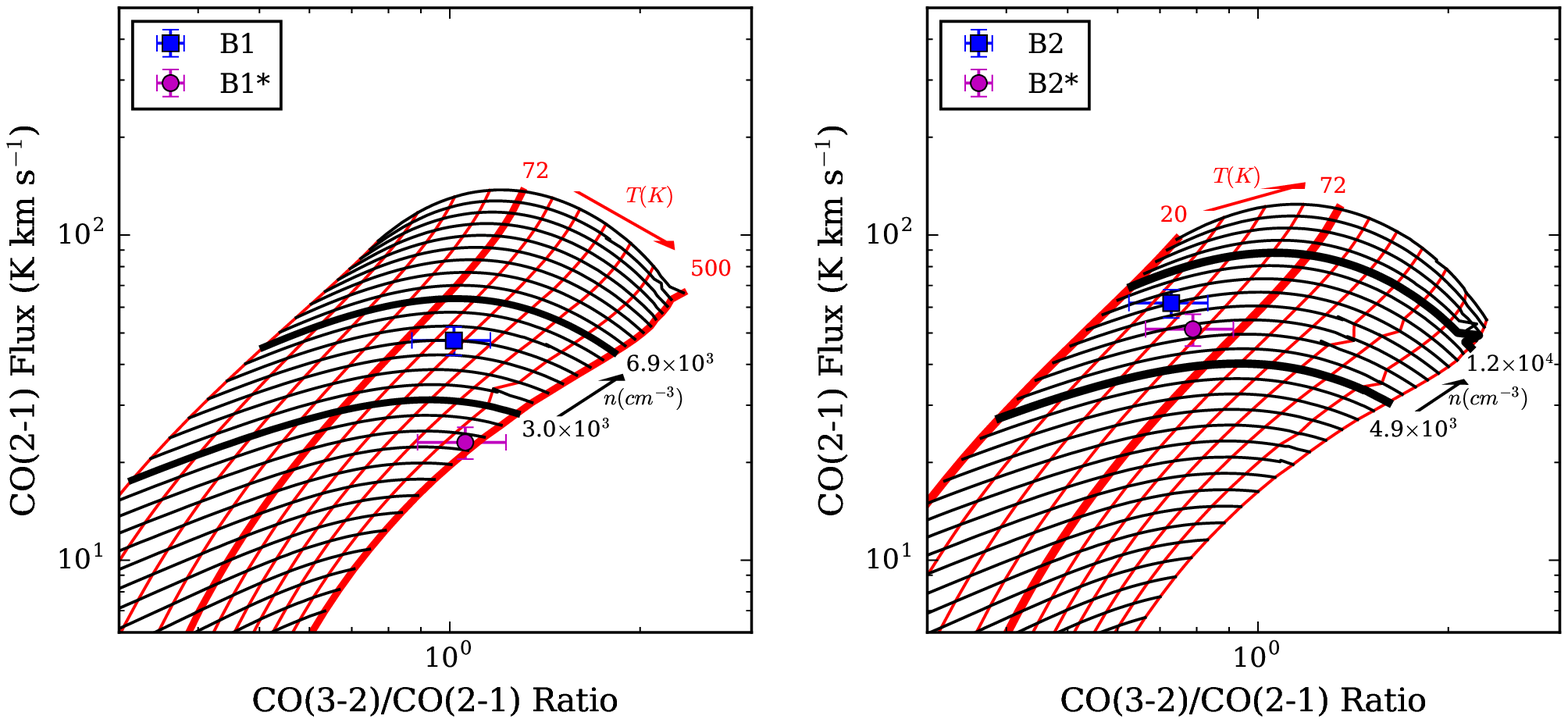}
\caption{ {\it Left:} LVG modeling results for bullet B1. Red and black lines denote models for a fixed temperature, $T$, and \hh~number density, $n$, respectively. Models with reference values are highlighted in bold. Data points with error bars show the observed CO 2--1 flux and the 3--2/2--1 ratio, with a filled blue square and magenta circle showing cases without and with the line wing component subtracted. {\it Right:} Same as the left panel, but for bullet B2.}
\label{fig:lvg}
\end{figure*}

\section{Discussions} \label{S:dis}
The two brightest bullets in the HH~80--81 SE outflow, B1 and B2, were first detected in the SMA CO 2--1 observations \citep{Qiu09a}. Here we present the SMA CO 3--2, SiO 5--4 and 8--7, \sob, and APEX CO 6--5 observations. The multi-tracer, multi-transition observations allow us to constrain the excitation conditions, chemistry, and origin of the bullets.

\subsection{Excitation conditions of the two bullets} \label{subS:excitation}
The CO 3--2 and 2--1 data, both of high angular resolutions and high SNRs, make it possible to quantitatively explore the gas temperature and density. The line ratio, CO 3--2/2--1, can be a relatively straightforward probe of the excitation conditions of molecular gas. B1 and B2 were simultaneously observed in the CO 2--1 or 3--2 line, thus whether the line ratio in B1 is greater or smaller than that in B2 is not affected by the flux calibration uncertainties, which are about 10\% in our SMA observations. The main issue that may affect the measurement of the line ratio is the separation of the flux levels between the bullets and a broad line wing (see Figure~\ref{fig:cospec}). Interestingly, the CO 3--2/2--1 ratio does not considerably change, measuring $\sim$1 in B1 and $\sim$0.7--0.8 in B2, for the cases with and without subtracting the line wing. Therefore the line ratio in B1 is higher than that in B2, no matter whether or not a broad line wing is subtracted, suggesting that the excitation conditions in B1 are higher than that in B2. 

We further explore the gas temperature and density of the bullets using the LVG calculations. The line ratio alone is not sufficient to constrain both $T$ and $n$. We then fit both the line ratio and the flux level in one of the two transitions. Since we are using the CO observations to probe the \hh~gas, there is another free parameter, $N_{\rm CO}/{\Delta}V$, in the LVG calculations. Thanks to a relatively stable and uniform CO-to-\hh~abundance ratio, we can convert $n$ to $N_{\rm CO}/{\Delta}V$ with a size scale which is derived by making a 2D Gaussian fitting to the deconvolved map. From the grid of LVG models shown in Figure~\ref{fig:lvg}, $T$ mainly depends on the line ratio while $n$ is more sensitive to the flux level. We then stress that the LVG fitting to $T$ is more reliable. On the other hand, $n$ is less stringently constrained, considering that the flux level of the bullets suffers from flux calibration uncertainties and the uncertainty in measuring the size scale, especially when the bullets are not fully resolved. 

In Figure~\ref{fig:lvg}, $T$ and $n$ change significantly from the case without subtracting the line wing to the case with the subtraction. The observations of the SiO and SO lines help us to determine which case is of more relevance. In Table \ref{tab:lines}, the SO and SiO lines have $n_{\rm cr}$ on orders of $10^6$--$10^7$~cm$^{-3}$; in particular, the SiO 8--7 line has $n_{\rm cr}\sim2\times10^7$~cm$^{-3}$. \citet{Shirley15} found that critical densities for commonly observed transitions tracing dense gas in molecular clouds are typically 1--2 orders of magnitude greater than effective excitation densities, which are defined as gas densities capable of producing a 1 K\kms~line. The SO and SiO lines are all detected in B1, with peak fluxes of $\gtrsim1$~K\kms~(see Figure \ref{fig:mole}), suggesting that B1 has a density of order at least $10^4$--$10^5$~cm$^{-3}$. However, the fitted density in B1 significantly decreases from $5\times10^3$ to $1\times10^3$~cm$^{-3}$ if the line wing component is subtracted. Moreover, the density in B1 is comparable to that in B2 for the case without subtraction, but becomes almost an order of magnitude lower than that in B2 for the case with the line wing subtraction. The lower density in B1 is inconsistent with the fact that the SO and SiO lines are detected in B1 but not in B2. Hence we speculate that the LVG calculations for the case without subtraction are more physical. In this case, the two bullets have comparable gas densities reaching $10^4$~cm$^{-3}$ while the temperature in B1 is about three times higher than that in B2. 

\begin{deluxetable*}{ccrr}
\tablecaption{Critical densities and upper level energies of the observed spectral lines \label{tab:lines}} 
\tablehead{ 
\colhead{Spectral line} & \colhead{Frequency} & \colhead{Critical density ($n_{\rm cr}$)\tablenotemark{a}} & \colhead{Upper level energy ($E_{\rm up}/k)$} \\
\colhead{} & \colhead{(GHz)} & \colhead{(cm$^{-3}$)} & \colhead{(K)}
}
\startdata 
CO 2--1  & 230.538 & (1.2--1.0)$\times10^4$ & 17 \\
CO 3--2  & 345.796 & (3.8--2.9)$\times10^4$ & 33 \\
CO 6--5  & 691.473 & (3.0--2.3)$\times10^5$ & 116 \\
\soa        & 219.949 & (2.2--2.1)$\times10^6$ & 35 \\
\sob        & 346.528 & (8.1--7.2)$\times10^6$ & 79 \\
SiO 5--4 & 217.105 & $4.7\times10^6$           & 31 \\
SiO 8--7 & 347.331 & $2.0\times10^7$           & 75 \\
\enddata
\tablecomments{Derived based on the Leiden Atomic and Molecular Database \citep{Schoier05}.}
\tablenotetext{a}{Calculated for gas temperatures ranging from 50 to 500 K for the CO, SiO lines and from 60 to 300 K for SO lines.}
\end{deluxetable*}

Excitation conditions of molecular bullets in low-mass outflows have been better studied. Estimates based on CO low level transitions ($J\leq4$) found a temperature $<150$~K and a density about $10^4$~cm$^{-3}$ \citep[e.g.,][]{Masson90,Chernin92,Cernicharo96,Hatchell99}. Multi-line observations of another molecule, SiO, suggested higher excitation conditions, with a temperature of a few 100~K and a density of $10^5$--$10^6$~cm$^{-3}$, when high level transitions (up to $J=11$--10) are included \citep[e.g.,][]{Nisini02,Nisini07}. Our LVG analyses of the CO 3--2 and 2--1 emissions in B1 and B2, which are associated with a high-mass protostar, obtain a temperature of a few 10 K to 100 K and a density reaching $10^4$~cm$^{-3}$, both in agreement with the CO observations of bullets in low-mass outflows. Multiple SO and SiO lines are also detected in B1, but the low SNRs in the high excitation lines (\sob~and SiO 8--7) do not allow a quantitative analysis. The CO 6--5 emissions are seen in both B1 and B2, but the low spatial resolutions and low SNRs hamper the inclusion of this line in our LVG calculations. It is possible that there exists higher excitation gas not well traced by the CO~3--2 and 2--1 emissions.

\subsection{On the origin of the bullets}
Bullets B1 and B2 are part of a collimated and EHV molecular outflow extending from the high-mass protostellar core, MM2, to about 0.2~pc to the southeast \citep{Qiu09a}. Molecular outflows are often interpreted as the product of interactions between fast moving flows, which are in the form of a collimated jet or a wide-angle wind (or a combination of both), and the ambient molecular clouds \citep{Richer92,Arce07}. Despite the long standing debates on the driving mechanisms of jets and winds, molecular outflows are implicitly a secondary effect in this picture. This interpretation is still relevant in particular for large scale ($\gtrsim0.1$~pc) outflows that often have masses significantly higher than those of the central forming star and move at relatively low velocities ($\sim1$--10\kms). It should be noted that ALMA observations have started to reveal small scale ($\lesssim100$~AU) molecular flows directly ejected from disks around both low-mass and high-mass young stellar objects (YSOs) \citep{Bjerkeli16,Tabone17,Hirota17}. On the other hand, extensive observations at optical, near-infrared, and centimeter to millimeter wavelengths show that EHV jets in low-mass YSOs are mainly molecular at the earliest Class 0 stages, and become more and more atomic from relatively late Class 0 to Class I and Class II stages \citep{Frank14,Bally16}. 

\begin{figure*}
\epsscale{1.}\plotone{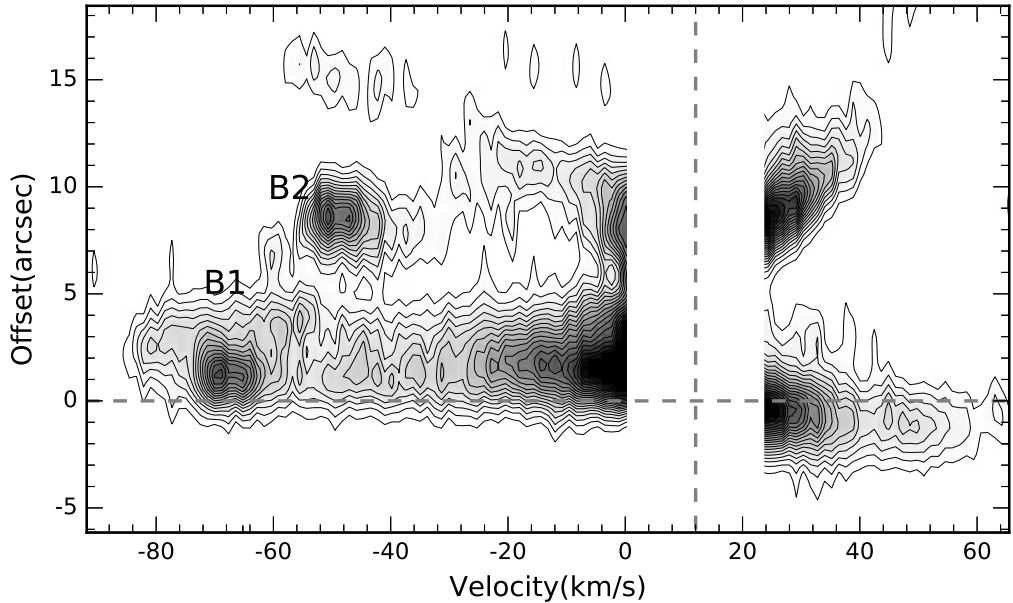}
\caption{CO 3--2 position-velocity diagram constructed along the jet axis as indicated by a grey line in Figure~\ref{fig:CO}; a vertical dashed line denotes the cloud velocity and a horizontal line indicates the peak position of MM2 \citep[see][]{Qiu09a}.}
\label{fig:pv}
\end{figure*}

What is the nature of bullets B1 and B2 in HH 80-81? Do they come from an ambient cloud, or originate from the close vicinity of the central protostar? The bullets are compact, with sizes of $\sim$0.01~pc, and move at the plane-of-sky velocities $>50$\kms. \citet{Qiu09a} suggested that the overlapping blue- and redshifted CO 2--1 emissions along the jet path may imply a small inclination angle of about 10$\arcdeg$, and thus the bullets velocity may be even higher, exceeding 100\kms~after correcting for the projection effect. A more accurate determination of the bullets velocity awaits future observations, which in combination with the existing data, should allow proper motion measurements of the bullets. Anyway, a velocity of order 100\kms~challenges the interpretation that the bullets are ambient materials accelerated by shocks, since in molecular clouds, hydrodynamic shocks with velocities $\geq$25\kms~are dissociative (J-type), and would destroy molecules \citep[e.g.,][]{Hollenbach77}, whereas non-dissociative (C-type) MHD shocks in molecular clouds can attain velocities no higher than 50\kms~\citep{Flower03}. Even in an unlikely scenario that the bullets contain ambient material gradually accelerated to extremely high velocities without introducing J-type shocks, there is a great difficulty for such a process to maintain the coherent structures so that the bullets remain compact and a small velocity dispersion. Efficient reformation of molecules in the post-shock gas \citep{Neufeld89,Hollenbach89} is probably the only mechanism compatible with the interpretation that the bullets consist of the ambient gas accelerated into motions by fast J-type shocks. In this scenario, the bullet material is presumably swept up by jet bow-shocks or dragged forward by a central jet through turbulent entrainment. However, in either case the bullets are expected to exhibit a ``Hubble-wedge'' patten, i.e., the maximum velocity increasing with the distance from the central source along the jet axis \citep[e.g.,][]{Stahler94,Smith97,Downes99}. In Figure~\ref{fig:pv}, both bullets appear as distinct structures spanning narrow ranges in velocity and in distance from the protostar. In addition, there appears to be no Hubble-wedge structure in the position-velocity (PV) diagram for either bullet. Therefore, the bulk bullet materials do not seem to come from the ambient cloud accelerated by either C-type or J-type shocks.  

Jets in the earliest Class 0 low-mass protostars have been found to have a predominantly molecular composition \citep{Bally16}. The dust core harboring the central source of the HH~80--81 SE outflow, MM2, has a temperature of only 32~K and its chemistry suggests a very early evolutionary stage for the central protostar \citep{Qiu09a}. In this sense, it appears likely that the bullets are part of a protostellar jet driven from a central engine around the forming star. If a jet varies in velocity, the fast jet material will catch up with a slow gas ejected earlier, forming a two-shock structure which is referred as an IWS that may appear as a bright emission knot \citep[e.g.,][]{Raga92,Masciadri02}. Since the gas inside an IWS is compressed by internal shocks, it is ejected sideways while traveling along the jet flow, and produces a sawtooth PV structure as seen in the Class 0 source IRAS 04166+2706 \citep{Santiago09,Wang14,Tafalla17}. In Figure~\ref{fig:pv}, there are no apparent sawtooth features in the two bullets, which argues against the IWS model. Alternatively, the bullets could be directly ejected from the central protostar-disk system \citep{Bachiller90b}, as a consequence of an episodic enhancement in mass loss, which in turn manifests an episodic mass accretion. Such a ``bullet'' interpretation is consistent with the profiles of the CO spectra (Figure~\ref{fig:cospec}) which show a steep drop on the highest velocities and a gradual decline toward the lower velocities, indicating that the majority of the gas is moving on the terminal velocities, as expected for a compact ejecta moving outward. Such a trend is also seen in Figure~\ref{fig:cospec}, where both bullets show brightness peaks at higher velocities. The ``bullet'' model can also naturally explain the velocity difference between the two bullets: B1 is relatively newly ejected since it is located closer to the central source, and thus has a greater velocity, whereas B2 has slowed down to some extent while traveling further out within the jet flow. The excitation difference between the bullets also appears to support the bullet model: B1 is breaking out of the dense core and could be shock-heated while interacting with the surrounding dense gas; B2 has traveled downstream into a more tenuous medium and has cooled down to a much lower temperature. On the other hand, in the IWS model, both bullets are tracing internal shocks, and from their sizes (or compactness, see Table~\ref{tab:measured}) and brightness (see Figure~\ref{fig:cospec}), there is no hint that the B2 shocks are much weaker than those of B1,  making it difficult to interprete the excitation/temperature difference between B1 and B2.

\begin{figure*}
\epsscale{.8}\plotone{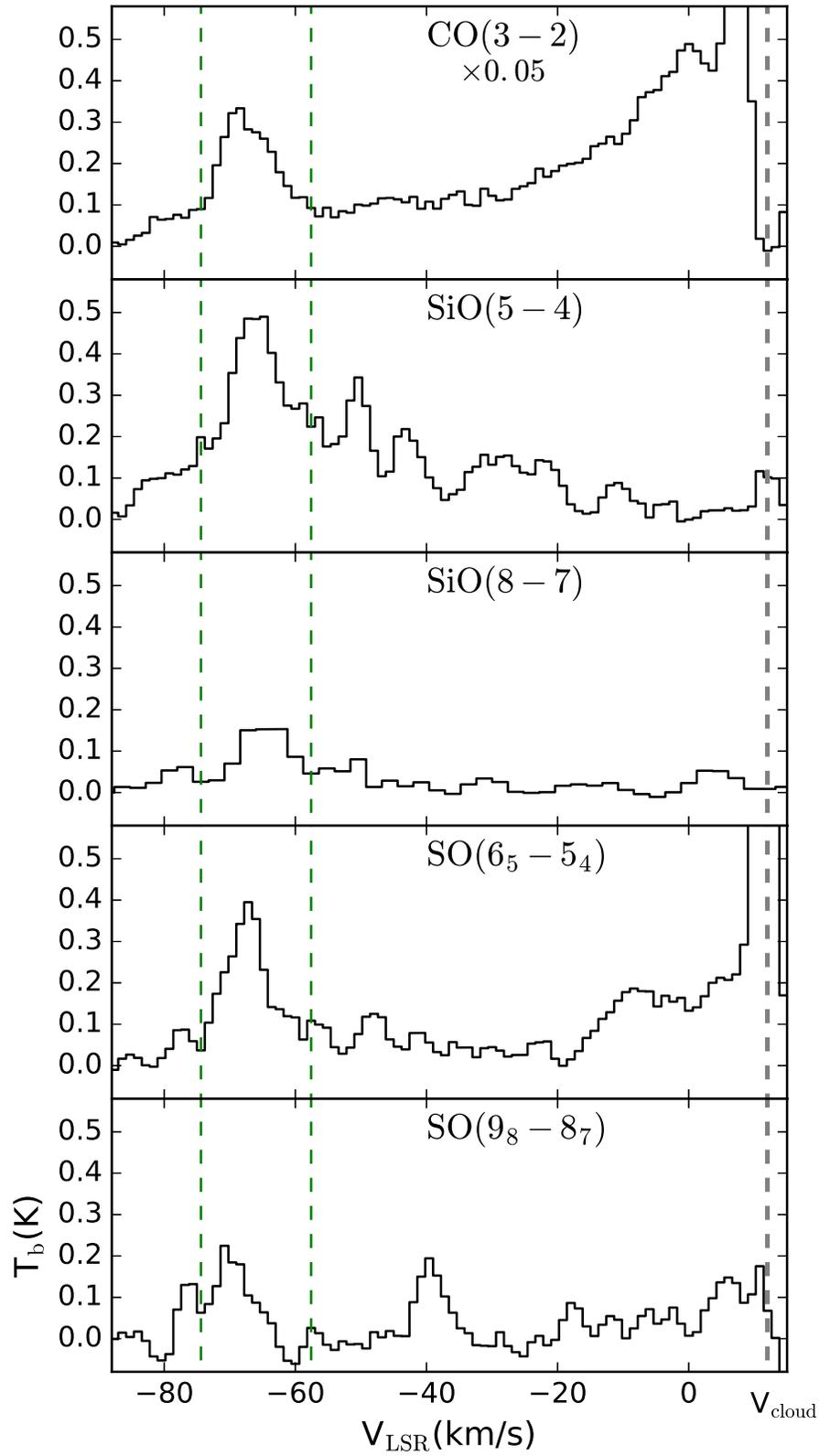}
\caption{CO 3--2, SiO 5--4, SiO 8--7, \soa~and \sob~spectra taken at the peak position of B1. The CO 3--2 brightness has been reduced by a factor of 20 for comparison. Two vertical dashed green lines indicate the velocity range of B1, and a vertical dashed gray line denotes the systemic velocity.}
\label{fig:mole}
\end{figure*}

The chemical abundance of an outflowing gas may provide key information on the physical origin of the gas. \citet{Tafalla10} performed a chemical survey of molecular outflows in two Class 0 sources. They found that the EHV gas is abundant in O-bearing molecules, such as SiO, SO, CH$_3$OH, and H$_2$CO, and is deficient in C-bearing molecules such as CS and HCN. They attributed this chemical peculiarity to an origin of the EHV gas in the innermost vicinity of the central protostar. The spectral coverage of our SMA and APEX observations includes CO, SO, and SiO lines, as well as a couple of CH$_3$OH and H$_2$CO lines. The latter are of higher excitations as compared to the CH$_3$OH and H$_2$CO transitions observed by \citet{Tafalla10}. We did not detect any CH$_3$OH or H$_2$CO emission in B1 or B2. No CS or HCN lines were covered by our observations. Nevertheless, a remarkable chemical difference between the two bullets is seen in our multi-line observations. The CO 2--1, 3--2, and 6--5 lines are all detected in both bullets, while the SO and SiO lines are only detected in B1 (Figure~\ref{fig:mole}, also see Figure~\ref{fig:int}). Could it be the excitation difference between the two bullets that causes this diversity? In Table~\ref{tab:lines}, the SO and SiO lines have critical densities 1--3 orders of magnitude higher than those of the CO lines, while the upper level energies are comparable among all the lines. The highest $E_{\rm up}/k$ occurs in the CO 6--5 transition (116~K) which is detected in both B1 and B2. Therefore the key physical condition required to excite the observed SO and SiO lines is a high density, rather than a high temperature. From our LVG calculations, B1 has a higher temperature but lower (or comparable) density than B2 (Section~\ref{S:lvg}), inconsistent with the possibility that the excitation conditions in B2 is too low to excite the SO and SiO lines. Instead the observations indicate that it is an chemical effect that leads to the detection of SO and SiO lines in B1 and the non-detection in B2. SO and SiO lines can both trace shocked gas in outflows \citep[e.g.,][]{Chernin94}. In particular, the SiO emission is a well known shock tracer in molecular clouds \citep{Schilke97}. If B1 and B2 are IWSs formed from internal shocks, provided that the two bullets show similar brightness and compactness, one would expect to detect SiO emissions in both bullets, such as the case of IRAS 04166+2706 \citep{Santiago09}. Thus the chemical difference between B1 and B2 cannot be easily accounted for in the IWS model. Again the ``bullet'' model provides a possible explanation that the newly ejected bullet B1 interacts with the dense core and creating shocks that give rise to the enhanced SiO abundance, whereas B2 travels in a less dense and fast moving flow such that no strong shock is induced. Such a scenario is also consistent with the SiO~5--4 and \soa~profiles which show additional peaks at lower velocities (Figure~\ref{fig:mole}), since multi-velocity components could be readily produced when B1 is interacting with the surrounding dense gas.

\section{Summary} \label{S:sum}
We present studies of two bright molecular bullets, B1 and B2, in high-mass star-forming region HH~80--81 in CO, SO, and SiO lines with the SMA and APEX. B1 is closer to the central source, and has a higher velocity and higher CO 3--2/2--1 line ratio than those of B2. An LVG analysis of the CO 2--1 and 3--2 lines yields gas temperatures of 70--210~K in B1 and 20--50~K in B2. Both bullets are detected in the CO 2--1, 3--2, and 6--5 transitions, whereas the SO and SiO emissions are only seen in B1, indicating that strong shock activities only take place in B1. The velocity structures of B1 and B2 strongly suggest that the bullets do not arise from accelerated ambient cloud gas. The PV diagram, excitation variation, and molecular chemistry all suggest that the two bullets arise from direct mass ejections from the innermost vicinity of the central protostar as a result of an episodic mass accretion in the formation of high-mass stars.

\acknowledgments K.Q. is supported by National Key R\&D Program of China No. 2017YFA0402600. K.Q. acknowledges supports from National Natural Science Foundation of China (Grant Nos. 11473011, U1731237, 11590781, and 11629302).

\facility{Submillimeter Array (SMA), Atacama Pathfinder EXperiment (APEX)}

\software{MIR (\url{https://github.com/qi-molecules/sma-mir}), MIRIAD \citep{Sault95}, RADEX code \citep{vdTak07}}


\end{document}